\documentclass[aps,prb,twocolumn, superscriptaddress]{revtex4-1}
\usepackage[latin2]{inputenc}
\usepackage{float}
\usepackage{graphicx}
\usepackage{hyperref}
\hypersetup{colorlinks=false}
\usepackage{amsmath}
\usepackage{xcolor}

\begin{document}

\title{Unsupervised feature recognition in single molecule break junction data}

\author{A.~Magyarkuti}
\affiliation{Department of Physics, Budapest University of Technology and Economics,\\ 1111 Budapest, Budafoki ut 8., Hungary}
\affiliation{MTA-BME Condensed Matter Research Group, Budafoki ut 8, 1111 Budapest, Hungary}

\author{N.~Balogh}
\affiliation{Department of Physics, Budapest University of Technology and Economics,\\ 1111 Budapest, Budafoki ut 8., Hungary}

\author{Z.~Balogh}
\affiliation{Department of Physics, Budapest University of Technology and Economics,\\ 1111 Budapest, Budafoki ut 8., Hungary}
\affiliation{MTA-BME Condensed Matter Research Group, Budafoki ut 8, 1111 Budapest, Hungary}

\author{L.~Venkataraman}
\affiliation{Department of Applied Physics, Columbia University, New York, New York 10027, USA}
\affiliation{Department of Chemistry, Columbia University, New York, New York 10027, USA}

\author{A.~Halbritter}
\affiliation{Department of Physics, Budapest University of Technology and Economics,\\ 1111 Budapest, Budafoki ut 8., Hungary}
\affiliation{MTA-BME Condensed Matter Research Group, Budafoki ut 8, 1111 Budapest, Hungary}

\date{\today}

\begin{abstract}
Single-molecule break junction measurements deliver a huge number of conductance vs.\ electrode separation traces. Along such measurements the target molecules may bind to the electrodes in different geometries, and the evolution and rupture of the single-molecule junction may also follow distinct  trajectories. The unraveling of the various typical trace classes is a prerequisite of the proper physical interpretation of the data. Here we exploit the efficient feature recognition properties of neural networks to automatically find the relevant trace classes. To eliminate the need for manually labeled training data we apply a combined method, which automatically selects training traces according to the extreme values of  principal component projections or some auxiliary measured quantities, and then the network captures the features of these characteristic traces, and generalizes its inference to the entire dataset. The use of a simple neural network structure also enables a direct insight to the decision making mechanism. We demonstrate that this combined machine learning method is efficient in the unsupervised recognition of unobvious, but highly relevant trace classes within low and room temperature gold--4,4' bipyridine--gold single molecule break junction data.
\end{abstract}

\maketitle

\section{Introduction}
Creating electrical circuit elements from single atoms or molecules is one of the main goals of molecular electronics research.\cite{Cuevas2010,Aradhya2013,Su2016a} The investigation of the electronic properties of single molecule junctions can be realized utilizing the break junction technique,\cite{Agrait2003,Xu2003} which allows us to establish a statistical amount of single-molecule nanowires by repeatedly opening and closing an atomic-sized metallic junction, and by appropriately dosing the target molecules. In spite of the stochastic nature of this process, the statistical analysis of the conductance vs. electrode separation traces usually yields typical, rather well defined single-molecule junction configurations, which are reflected by peaks in the conductance histograms. However, to understand the fine details of the typical junction formation trajectories, it is not enough to plot simple \emph{one dimensional} conductance histograms, rather some more elaborate techniques are required. These include two-dimensional conductance-displacement histograms,\cite{Martin2008a, Kamenetska2009} cross-correlation analysis\cite{Halbritter2010, Makk2012, Aradhya2013a, Magyarkuti2016} or custom feature filtering algorithms,\cite{Halbritter2003, Ulrich2006, Quek2007b, Makk2012a, Widawsky2013, Balogh2014, Dell2015, Su2015, Balogh2015, Adak2015a, Huang2017} which are able to identify targeted motifs of the traces. Although these methods rely on well defined computer algorithms instead of manual data selection, the feature filtering protocols are usually custom constructed and tuned according to the researcher's physical intuition.

Nowadays artificial intelligence methods are widely utilized in many fields of science and technology providing a rapidly developing tool to recognize the relevant features in the data without the guidance by human intuition. In molecular electronics it was also demonstrated that machine learning protocols, like unsupervised vector-based classification,\cite{Lemmer2016} reference free clustering method,\cite{Cabosart2019} fast data sorting with principle component analysis,\cite{Hamill2018} and neural network-based classification\cite{Lauritzen2018} may become useful tools for data analysis. In the latter work we have demonstrated the successful classification of single-atom and single-molecule break junction data relying on recurrent neural networks that were trained either on computer simulated data, or on manually selected and labeled experimental traces. Though being successful in the classification of the races, it is clear that this approach can be further optimized in various aspects. On one hand the rather complex recurrent neural networks were found to be sensitive to the choice of the network parameters, excellent classification was only achieved at some specific parameter sets. On the other hand the training method also requires improvement: whereas training on computer simulated data is expected to have increasing importance with the development of the simulations' predictive power, in case of experimental training sets it would be definitely favorable to eliminate the need for manual labeling. In this paper we step forward towards an unsupervised learning protocol. We apply a combined method using the simplest possible feed-forward neural network with a single hidden layer for feature identification. The training is either based on the principle component (PC) projection of the conductance traces, or on an additional measured quantity, like force. In both cases the two sides of the distributions (i.e. the extreme PC projections or rupture force values) are used for the training such that the network first learns from traces with clear features, and then generalizes for traces with less obvious characters. We demonstrate that this approach exhibits excellent performance on break junction data: (i) the classification results are insensitive to the precise choice of the network parameters; (ii) thanks to the simple network structure one can extract valuable information about the decision making mechanism; (iii) this unsupervised training protocol provides similar classification results as the human feature recognition. 

\section{Results and Discussion}
\begin{figure}
\begin{center}
\includegraphics[width=0.7\columnwidth]{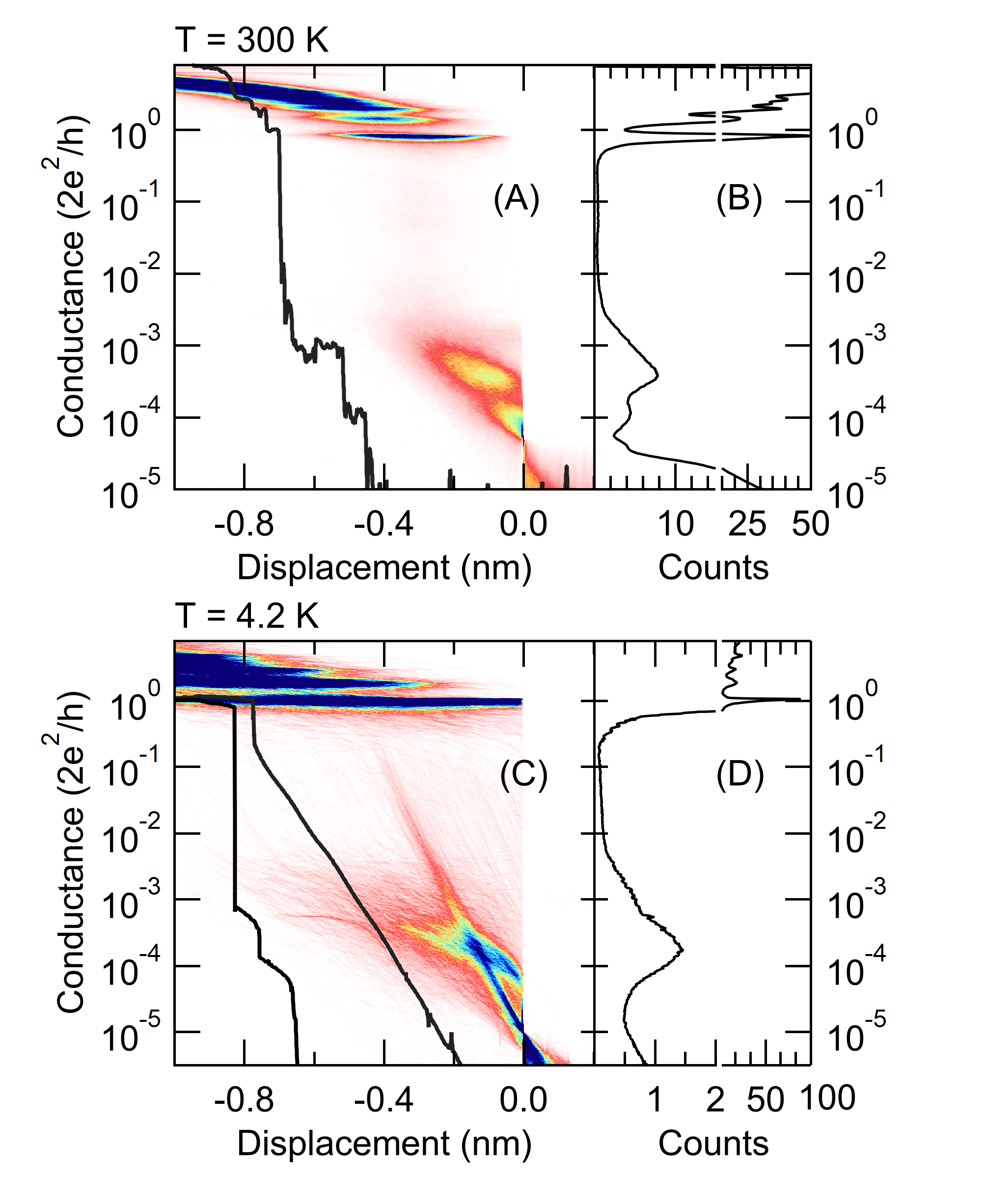}
\caption{\textit{
2D conductance vs.\ electrode separation histograms (left) and 1D conductance histograms (right) of gold-4,4' bipyridine-gold single-molecule junctions measured at room temperature (A, B) and at liquid helium temperature (C, D). The histograms of the room temperature dataset are constructed from $12000$ traces, while the low temperature dataset consist of $5500$ traces. In the 2D histograms the traces are aligned at $G_\textrm{ref}=5\cdot10^{-5}\,$G$_0$ (A) and $G_\textrm{ref}=10^{-5}\,$G$_0$ (C) conductance. Example traces from both datasets are shown as black lines (A, C). In panel (C) the left/right curves illustrate a double step molecular trace and a tunneling trace, respectively.}}
\label{Fig1}
\end{center}
\end{figure}

We analyze the conductance traces of gold-4,4' bipyridine (BP)-gold single molecule junctions that were acquired either by a room temperature scanning tunneling microscope break junction setup extended with force measurement \cite{Frei2011,Aradhya2012} or by a mechanically controllable break junction arrangement\cite{Makk2012a} operated at cryogenic temperatures ($T=4.2\,$K). In the former case, BP molecules are evaporated onto a gold coated mica substrate, whereas in the latter setup an in-situ evaporation method is applied.\cite{Magyarkuti2016}
It was shown by previous room temperature measurements that BP molecules can attach to gold electrodes in two different binding geometries resulting in double step molecular plateaus\cite{Quek2009,Hong2012,Aradhya2012a} (see the sample trace in Fig.~\ref{Fig1}A). At a smaller electrode separation the molecule can bind on the side of the metallic junction such that both the nitrogen linker and the aromatic ring is electronically coupled to the electrode. Upon further pulling, the molecule slides to the apex and only the linkers couple to the electrodes yielding a decreased conductance value. In the following we refer to these two binding geometries as HighG and LowG configurations.  
In accordance with these configurations the room temperature 1D conductance histogram exhibits double peaks (Fig.~\ref{Fig1}B), and similarly the 2D conductance-electrode separation histogram exhibits two plateaus (Fig.~\ref{Fig1}A). It was also found that the molecule pick-up rate is $\approx100\%$, i.e. almost all the conductance traces contain molecular plateaus. Note, that the single-molecule junction forms after the rupture of a single-atom gold nanowire, and the latter is reflected by a sharp peak in the conductance histogram (Fig.~\ref{Fig1}B) around the quantum conductance unit ($G=1\,$G$_0=2e^2/h$).

The cryogenic temperature measurements exhibit three clear differences compared to the room temperature data: (i) the 1D histogram exhibits a single peak around the LowG region (Fig.~\ref{Fig1}D); (ii) the pick-up rate significantly decreases ($\approx 30-40\%$); (iii) the stability of the junction is significantly increased.  The latter two features are clearly reflected by the 2D histogram: due to the reduced pick-up rate the traces with molecular plateaus are mixed with tunneling traces where the exponential decay of the tunnel current between the metallic apexes is clearly resolved due to the enhanced stability (see the 2D histogram and a sample molecular and tunneling trace in Fig.~\ref{Fig1}C). 

\begin{figure}
\begin{center}
\includegraphics[width=\columnwidth]{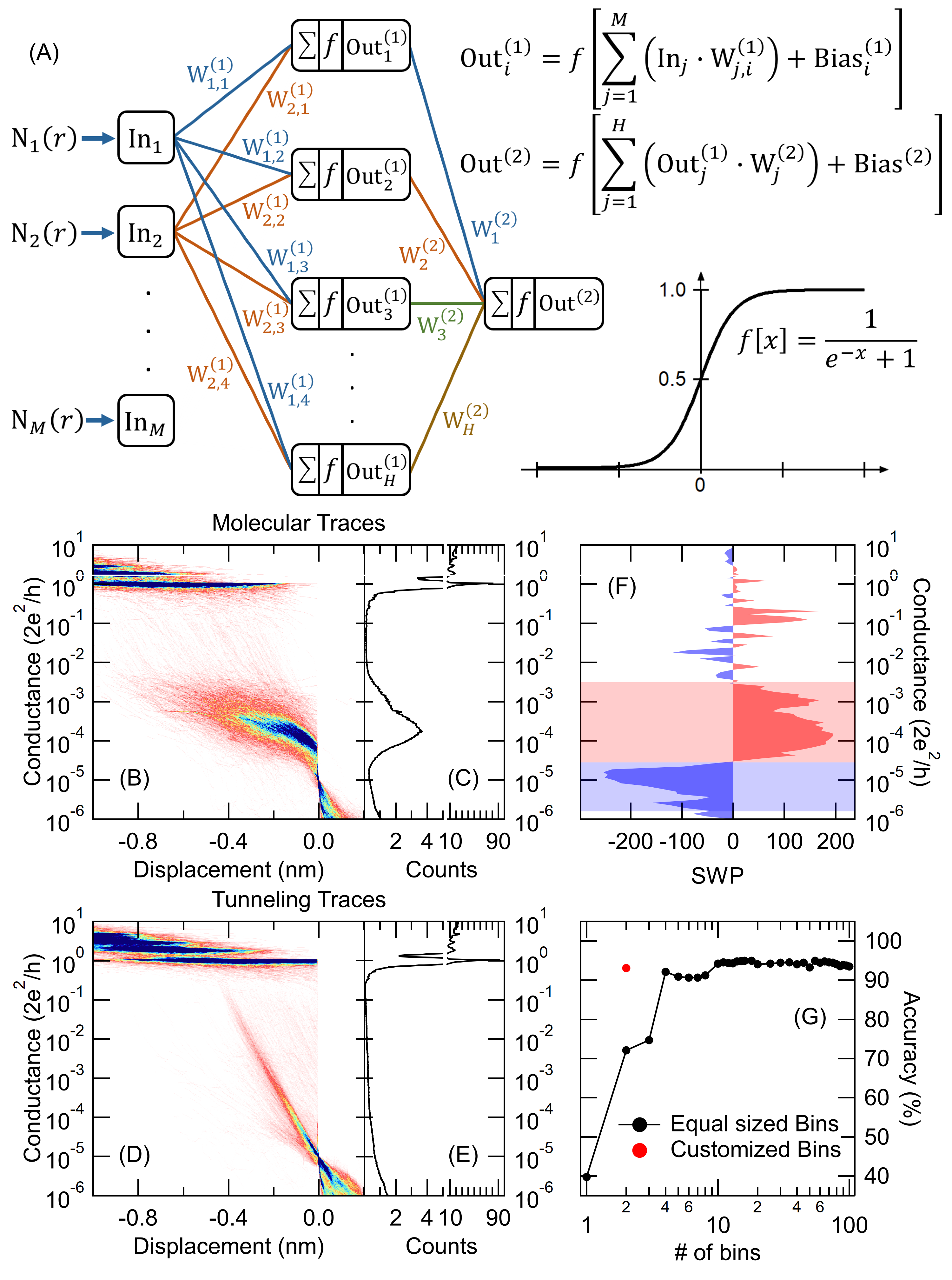}
\caption{\textit{(A) Schematic representation of the applied feed-forward neural network with one hidden layer and a single neuron at the output. The network takes the 1D conductance histograms of the individual conductance traces ($N_i(r)$) as inputs. Each neuron in the network sums up the incoming signals multiplied by the weights ($W_{j,i}$), adds a bias offset and applies a sigmoid activation function (f[x] shown in the inset). After the training (optimization of the weights and bias values of the network), the neural network classifies all traces measured at low temperature either as molecular or tunneling trace. Panels (B,C) / (D,E) show the 2D and 1D histograms of the such classified molecular/tunneling traces, respectively.
In the 2D histograms the traces are aligned at $G_\textrm{ref}=10^{-5}\,$G$_0$. Panel (F) shows the summed weight products of the trained neural network as a function of the conductance (see text). Panel (G) illustrates the network's classification accuracy as the function of the input bin number, $M$. The red dot shows the accuracy using the two customized bins extracted from the SWP plot in panel (F).}}
\label{Fig2}
\end{center}
\end{figure}

In the following we analyze these datasets using the neural network illustrated in Fig.~\ref{Fig2}A. The $\textrm{In}_i$ input vectors of the feed-forward neural network are simply the histograms of the individual conductance traces, $N_i(r)$, the number of data points in bin $i$ on trace $r$. This histogram is restricted to the $G=10^{-5}-10^{1}\,$G$_0$ conductance range for the room temperature measurement and $G=10^{-6}-10^{1}\,$G$_0$ in the case of the low temperature measurement using logarithmic binning. The size of the network's input vector, and accordingly the $M$ number of bins in the histogram is an adjustable parameter of the network together with the number of neurons in the hidden layer ($H$). The neurons in the hidden layer sum up the incoming signals using the weights of the synapses between the input and the hidden layer and placing a bias offset. Finally, these neurons output the summed signals applying a nonlinear (sigmoid) activation function. These outputs of the hidden layer are similarly fed to the output layer with a single neuron. The output value can be interpreted as a result of binary classification, e.g. the trace is classified as molecular/tunneling trace if the network output is larger/smaller than 0.5.
The network is trained on a subset of the traces that are labeled according to a specific criterion (e.g. molecular trace vs. tunneling trace). Along the training process the optimized weight and bias values are found using the backpropagation algorithm implemented in the TensorFlow machine learning platform.\cite{Abadi2016} Finally, the trained network is ready to classify any conductance trace, also those that were not used for training.

In this paper, we demonstrate the classification performance of these neural networks using the following scenario: (i) As an initial test we separate molecular traces from tunneling traces using the cryogenic temperature data. Relying on the manual labeling of all traces we first train the network on a part of the traces and then we evaluate the classification accuracy on the entire dataset. Finally we study the decision making mechanism by analyzing the weight-products of the network. (ii) We perform the same classification by replacing the manually labeled data with a training set that is automatically generated by principle component projection. (iii) We further analyze the molecular traces applying an additional principle component decomposition on the molecular traces classified in (ii). The results of this analysis explain the difference between the room and low temperature conductance histograms. (iv) Finally, we analyze the room temperature traces, where the conductance data are supplemented with force measurements. The force data are only used to label the training traces, demonstrating that afterwards the network is able to identify the relevant trace classes using solely the conductance data. The results of these classification tasks demonstrate, that our protocol performs well in automatically finding the relevant parts of the traces, reflecting distinct junction formation trajectories. This would be a challenging and time consuming task by manual data analysis, and therefore our combined classification method provides a valuable guidance in understanding the physical processes in single-molecule junctions. 

\emph{(i) Classification of molecular traces against tunneling traces using manually labelled training sets.}
This classification task serves as a reference to test the neural network's performance. We have manually labeled all the traces (5500 curves) using three categories: tunneling trace ($59\%$), molecular trace ($34\%$) and ill-defined trace ($7\%$). The \emph{accuracy} of the classification is measured as the [number of well characterized tunneling + molecular traces]/[number of all molecular and tunneling traces]. In this first approach we train the neural network using a subset of the data containing 320 traces with clear molecular or tunneling characteristic. Afterwards the neural network classifies all traces, as demonstrated by the 1D and 2D histograms in Fig.~\ref{Fig2}B,C (molecular traces) and Fig.~\ref{Fig2}D,E (tunneling traces). These figures do not exhibit any sign of misclassification, no molecular/tunneling signature is visible on the 2D histogram of the tunneling/molecular traces, respectively. Relying on the manual labeling of all traces, the network achieves $93\%$ classification accuracy. 

It is important to note, that the manual labeling of most traces is obvious, but it is a more delicate task to define a custom filtering algorithm for that. The most obvious filtering would rely on the number of datapoints within the conductance region of the molecular plateau assigning molecular/tunneling label to the traces with larger/smaller number of points than a proper threshold. This algorithm provides significantly worse classification accuracy ($\approx85-90\%$ depending on the chosen conductance range) than the neural network. Furthermore, this simplified method systematically misclassifies around $25-30\%$ of the molecular traces, whereas the neural network algorithm provides a misclassification rate below $10\%$. From this comparison it is clear that such a simple criterion is insufficient for the proper classification,  rather some combined features should be measured, including e.g. the slope of the trace within a proper region, or comparing the number of datapoints in multiple conductance regions. 

The neural network algorithm automatically finds a proper combined classification criterion. Due to the simple structure of our neural network it is also possible to get a quantitative insight to the networks's decision making algorithm. As a simplest measure, we can calculate the summed weight products for all the routes between a certain input and the output, 
$\textrm{SWP}_i=\sum_{j=1}^{H}{W_{i,j}^{(1)}\cdot W_j^{(2)}}$. If $\textrm{SWP}_i$ is a large positive/negative number for a certain input, a large input value (i.e. a large histogram count) will push the decision towards the molecular/tunneling label, respectively. If $\textrm{SWP}_i$ is close to zero for a certain input, then this input is not relevant in the decision making process. For our particular network, the SWP plot displays large positive values in the region of the molecular plateau (see the region with light red background in Fig.~\ref{Fig2}F), and a large negative region is observed at lower conductances (see the region with light blue background in Fig.~\ref{Fig2}F). In the latter region the molecular traces display a jump, but the tunneling traces contain significant counts. It is clear from the SWP plot that the network checks combined criteria: to give a trace a tunneling label it is not enough to have a small number of points in the region of the molecular peak; there should be enough points at even lower conductance values, where a molecular trace would exhibit a jump. Checking combined criteria brings clear improvement in the classification of molecular traces compared to the above misclassification of the molecular traces using a single criterion.

For the above analysis we have used 100 input neurons, i.e. 100 histogram bins. Next we check the stability of the neural network performance against the number of the input bins ($M$). Fig.~\ref{Fig2}G demonstrates, that the reduction of the bin number even slightly increases the classification accuracy down to $\approx10$ bins, but below that the accuracy drops. However, once we know the relevant conductance regions for the decision making from the SWP plot, we can define customized bins to focus our analysis on the most relevant regions. In this particular case we can reduce the number of input neurons to two, by calculating the number of datapoints in the two relevant regions of the SWP plot (i.e. the regions with light red and light blue background). This highly simplified network also achieves $93\%$ accuracy (see the red dot in Fig.~\ref{Fig2}G). We generally use $H/M=1.5$ ratio, where H is the number of neurons on the hidden layer but a broader region around this value also provides similar classification results.   

\begin{figure}
\begin{center}
\includegraphics[width=\columnwidth]{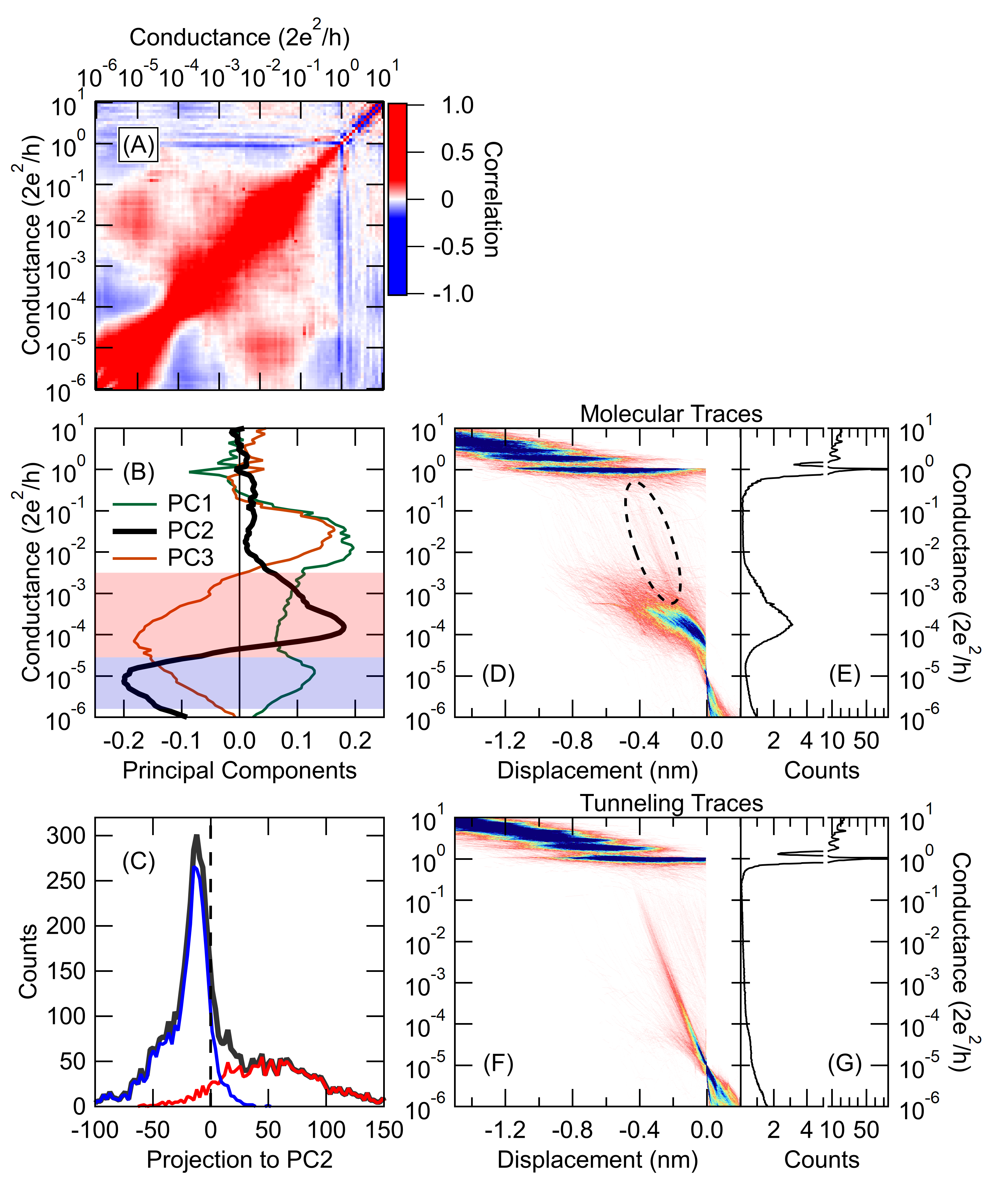}
\caption{\textit{
Classification of the low temperature tunneling/molecular traces using principal component projections. (A) The $C_{i,j}=\left\langle \delta N_i(r) \cdot \delta N_j(r) \right\rangle_r/\sqrt{\left\langle [\delta N_i(r)]^2\right\rangle_r\left\langle [ \delta N_j(r) ]^2 \right\rangle_r}$ correlation plot of the entire dataset, where $i,j$ represent the conductance bin labels, $\delta N_i(r)=N_i(r)-\left\langle N_i(r) \right\rangle_r$, and the $\left\langle\ \right\rangle_r$ averaging is performed along the $r$ trace index. (B) Principal components of the correlation matrix corresponding to the three largest eigenvalues. The light red/blue regions are reproduced from the SWP plot in Fig.~\ref{Fig2}F as a reference. (C): Distribution of the PC2 projections for all measured traces (black), manually labeled molecular (red) and tunnelling (blue) traces. Traces with positive/negative projection are classified as molecular/tunneling trace. Conductance histograms of the such classified molecular (D, E) and tunneling (F, G) traces. The encircled region in panel (D) illustrates that a significant portion of the tunneling traces are misclassified. In the 2D histograms the traces are aligned at $G_\textrm{ref}=10^{-5}\,$G$_0$.
}}
\label{Fig3}
\end{center}
\end{figure}

\emph{(ii) Unsupervised classification of molecular and tunneling traces using the combination of principal component decomposition and neural network analysis.}
Next we train our network without a manually labeled dataset. Manual classification is not only against objective data handling, but in many cases we also lack the a priori knowledge for judgement, and therefore we seek computer algorithms to automatically find the relevant data classes, which would help us to understand the various possible junction configurations. To this end we apply the method of principle component analysis developed by J. Hamill at al. \cite{Hamill2018}. This method relies on the correlation analysis of the conductance traces introduced in our previous work \cite{Makk2012}. The principal components (PCs) are the eigenvectors of the 2D correlation matrix, the PCs with the largest eigenvalues identify the most relevant correlations in the dataset.  Fig.~\ref{Fig3}A demonstrates the 2D correlation plot for our low temperature BP data, whereas Fig.~\ref{Fig3}B shows the principle components with the first three largest eigenvalues. The second principal component (PC2 shown by thick black line) exhibits a very similar shape to the SWP plot in Fig.~\ref{Fig2}F (for a better comparison the light red/blue regions of the SWP plot are repeated in Fig.~\ref{Fig3}B). This suggests, that the projection of the conductance traces to PC2 (i.e. the $\sum_{i=1}^M{PC2_i\cdot N_i(r)}$ scalar product) is able to classify molecular traces against tunneling ones. Indeed, the principal component analysis can alone perform the classification task giving molecular/tunneling label to the traces with positive/negative principal component projection. The 1D and 2D histograms for the such classified traces are shown in Fig.~\ref{Fig3}D-G. This classification provides a basically good result with $90\%$ accuracy, but the 2D histogram of the traces with molecular label (Fig.~\ref{Fig3}D) demonstrates that a significant amount of the tunneling traces are misclassifed as molecular ones (see the encircled region). The reason for this becomes clear if the histogram of the principal component projections (black line in Fig.~\ref{Fig3}C) is decomposed according to the manual labelling of our traces (see the red/blue lines in Fig.~\ref{Fig3}C for the molecular/tunneling traces, respectively). The zero PC projection (black dashed line) is clearly not the proper threshold between the two classes, as it cuts the distribution at a region where the tunneling traces are still highly dominant (at this line $76\%$ of the traces is tunneling trace, which is well above the overall $59\%$ ratio of the tunneling traces). Due to this improper threshold $11\%$ of the tunneling traces are misclassified. Alternatively, one could set the classification threshold at the maximum of the the PC projections' distribution, but this choice would further increase the number of misclassified traces, and reduce the classification accuracy. Without the a priori knowledge of the trace labels one does not have any basis to determine a better threshold.

\begin{figure}
\begin{center}
\includegraphics[width=\columnwidth]{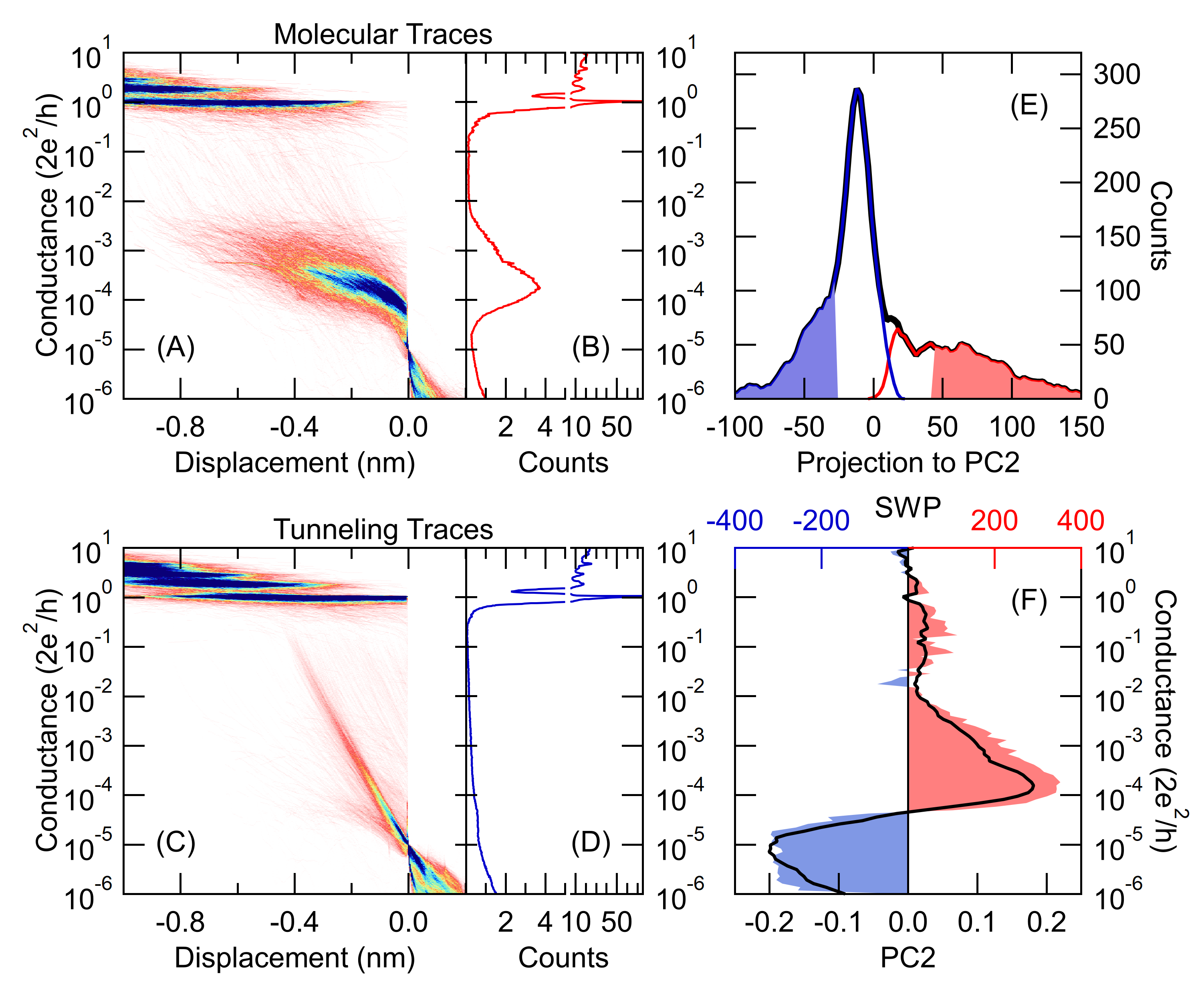}
\caption{\textit{
Combined principal component and neural network method for sorting the traces measured at low temperature. 2D and 1D conductance histograms of the traces classified as molecular (A, B) and tunneling (C, D) trace. In the 2D histograms the traces are aligned at $G_\textrm{ref}=10^{-5}\,$G$_0$. (E) Distribution of the PC2 projections for all measured traces (black), training traces labeled as molecular/tunneling trace (light red/blue area), traces classified as molecular/tunneling trace (red/blue line). (F) SWP plot of the trained neural network (red and blue area). As a reference, PC2 of the correlation matrix is reproduced from Fig.~\ref{Fig3}B (black line).}}
\label{Fig4}
\end{center}
\end{figure}

To solve the above problem of simple PC analysis, we apply a combined approach. We first take the traces from the two sides of the principle component projections' distribution (see light red and light blue regions in Fig.~\ref{Fig4}E, where both regions include $20\%$ of all traces). These traces clearly exhibit the features of the two classes showing definite tunneling/molecular characters, therefore these two trace sets provide an ideal training set for the neural network illustrated in Fig.~\ref{Fig2}A. During the training, the neural network learns the relevant features of these two trace classes, and then it generalizes these features for the rest of the traces with less clear character. This combined classification not only resolves the indefinite threshold problem of the principal component projections, but the neural network may also recognize more sophisticated features, which could not be captured by a simple principle component analysis. Performing this combined analysis we achieve $93\%$ classification accuracy, and the ratio of misclassified tunneling traces is reduced below $3\%$. The 2D and 1D histograms of the corresponding traces labeled as molecular/tunneling curves are demonstrated in Fig.~\ref{Fig4}A,B and Fig.~\ref{Fig4}C,D respectively. Both the 1D and 2D histograms confirm, that the misclassification of a significant amount of traces is avoided with this analysis. The SWP figure (Fig.~\ref{Fig4}F) exhibits a similar structure as the SWP plot in our previous analysis using the manually labeled training set (Fig.~\ref{Fig2}F).

This analysis demonstrated, that the efficient unsupervised feature recognition is clearly a mixed effort of principal component and neural network analysis: the PC projections are able to deliver a proper, automatically generated training set, but without neural network supplement the PC analysis would miss the proper classification thresholds.

\begin{figure}
\begin{center}
\includegraphics[width=\columnwidth]{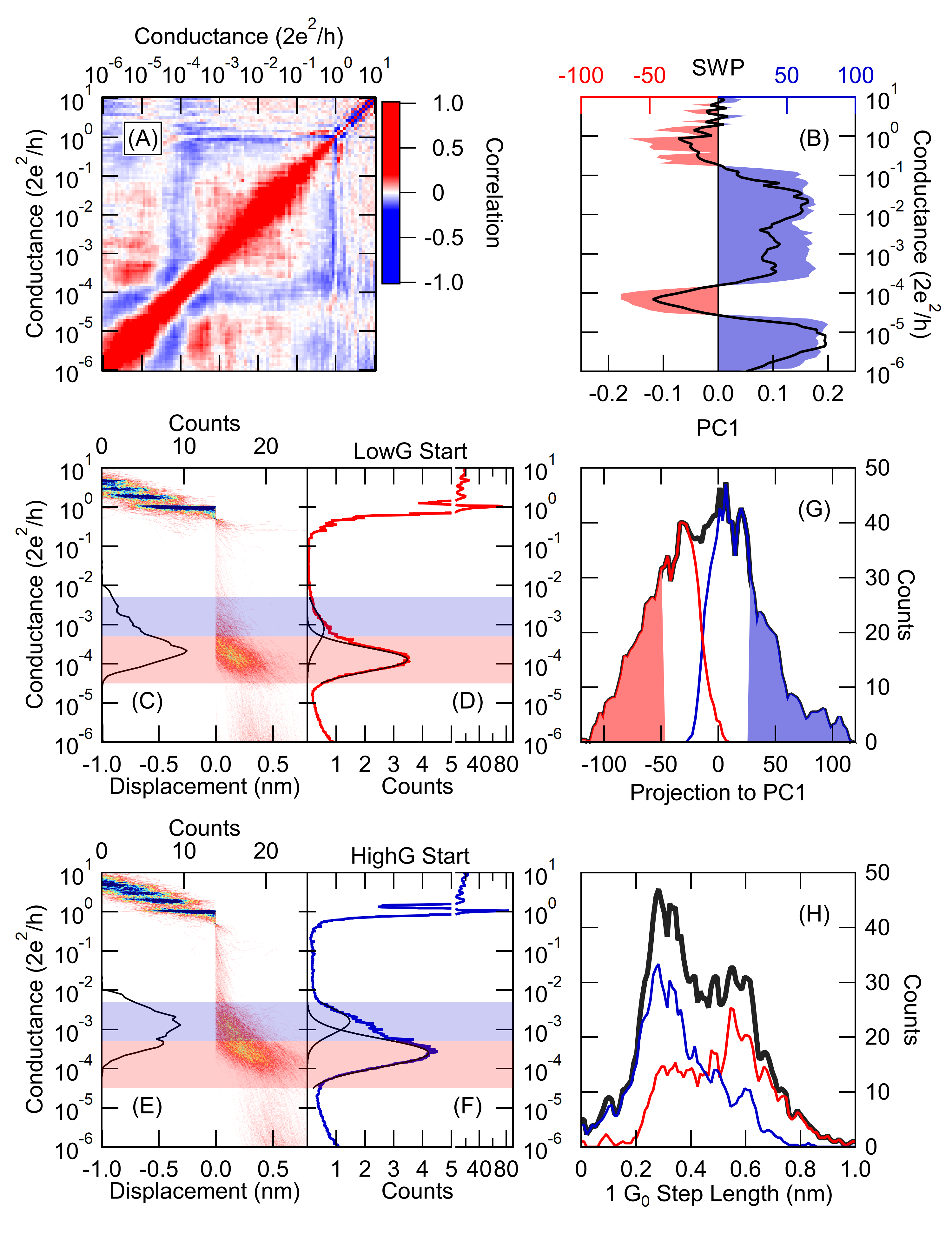}
\caption{\textit{
Unsupervised identification of the relevant trace classes among the molecular traces (i.e. using the traces with molecular label according to Figs.~\ref{Fig4}A,B) (A) correlation matrix of the traces with molecular label. (B) PC1 (black): principal component corresponding to the largest eigenvalue. SWP plot of the trained neural network (red and blue area). (C-F) 2D and 1D conductance histograms of the resulting two trace classes.  To highlight the initial part of the traces $G_\textrm{ref}=0.5\,$G$_0$ alignment is applied in the 2D histograms. The initial configuration histograms of the two trace classes are shown by black lines in panels (C) and (E), respectively. For the first, LowGStart trace class (C,D) the molecular plateau starts in the low conductance region (light red region), wheres in the second, HighGStart trace class (E,F) the molecular plateau starts in the high conductance region (light blue region). (G) Distribution of the PC1 projections: all molecular traces (black line), traces used for training (light red and blue area), resulting trace classes: LowGStart (red line), HighGStart (blue line). (H) Step length distribution of the $1G_0$ plateau: all molecular traces (black line), LowGStart traces (red line), HighGStart traces (blue line).
}}
\label{Fig5}
\end{center}
\end{figure}

\emph{(iii) Unsupervised recognition of distinct molecular trace classes.}
As a further step we take the traces classified as molecular curves according to our combined principal component and neural network analysis (see Figs.~\ref{Fig4}A,B). On this restricted trace set (already excluding the tunneling traces) we apply an additional principal component analysis relying on the first principle component (thick black line in Fig.~\ref{Fig5}B) of the correlation matrix in Fig.~\ref{Fig5}A. Again, we use the two sides of the principal component projections' distribution as the training set for our neural network (see light red and light blue regions in Fig.~\ref{Fig5}G including $20-20\%$ of all traces). The SWP figure of the trained network (area graph in Fig.~\ref{Fig5}B) shows similar structure as the first principal component (thick black line). The results of the classification are illustrated in Fig.~\ref{Fig5}C-F as 1D and 2D histograms. These two trace classes exhibit a very clear difference: whereas the traces of the first class (Fig.~\ref{Fig5}C,D) dominantly start from the LowG region (light red area in the 2D histogram), the traces of the second class (Fig.~\ref{Fig5}E,F) are rather starting in the HighG region (light blue area in the 2D histogram) and they only reach the LowG region after further stretching. This clear difference is also confirmed by the initial configuration histograms, i.e. the histograms relying on the datapoints within the first $0.1\,$nm displacement after the rupture of the gold nanowire (see the black histograms in Fig.~\ref{Fig5}C,E). According to this distinction, in the following we refer to these two trace classes as LowGStart and HighGStart traces. For both trace sets we fit the 1D histogram with double Gaussians (see black lines in Fig.~\ref{Fig5}D,F). For the LowGStart traces the 1D histogram (Fig.~\ref{Fig5}D) is clearly dominated by a single gaussian positioned in the LowG region. On the other hand both gaussian contributions are significant in the conductance histogram of the HighGStart traces (Fig.~\ref{Fig5}F). Accordingly, the HighGStart traces resemble the room temperature measurements on gold-BP-gold junctions reflecting double molecular plateaus in the HighG and LowG regions. In contrast the LowGStart traces represent a trace class, which is not common in room temperature data. 

Both PC1 and the SWP plot exhibit a further remarkable phenomenon showing large negative weights in the region of the $1\,$G$_0=2e^2/h$ quantum conductance unit (Fig.~\ref{Fig5}B). This means, that a long single-atom plateau with $\approx 1\,$G$_0$ conductance would push the classification towards the LowGStart label. To test this consideration we plot the length distribution of the $1\,$G$_0$ plateaus (i.e. the step length histogram shown by black line in Fig.~\ref{Fig5}H), which displays double peaks. This is a clear indicator of monoatomic chain formation.\cite{Yanson1998} After decomposing the step length histogram according to the two trace classes, it becomes clear that the second step length histogram peak is suppressed for the HighGStart traces, which means that these traces dominantly appear if the gold monoatomic contact breaks without chain formation. For the LowGStart traces, however, the first step length histogram peak is suppressed, and the second peak is enhanced. This means, that these traces dominantly appear if a monoatomic chain was already pulled before the rupture of the gold wire. In the latter case the chain atoms relax back to the electrodes after the rupture leaving a significantly larger gap between the apexes than in the former case, when the gold junction breaks without chain formation.  In case of the HighG molecular configuration the aromatic ring also binds to the side of the electrodes, but such a configuration cannot accommodate larger gaps, i.e. after the rupture of monoatomic gold chains this configuration is typically missing. Due to the enhanced mechanical stability at cryogenic temperatures a sufficiently large portion of the traces exhibit atomic chain formation, which also brings the clear dominance of the LowG molecular peak in the low temperature 1D conductance histogram in contrast to the dominance of the HighG peak at room temperature.

This analysis demonstrated that our combined classification algorithm automatically found the two relevant trace classes of LowGStart and HighGStart traces, and the structure of the principal component/SWP plot gave us a relevant hint that the $1\,$G$_0$ step length acts as a precursor of the molecular trace classes. These conclusions did not require any prior knowledge about the dataset; the classification algorithm recognized the relevant motifs of the traces in an unsupervised way.

\begin{figure}
\begin{center}
\includegraphics[width=\columnwidth]{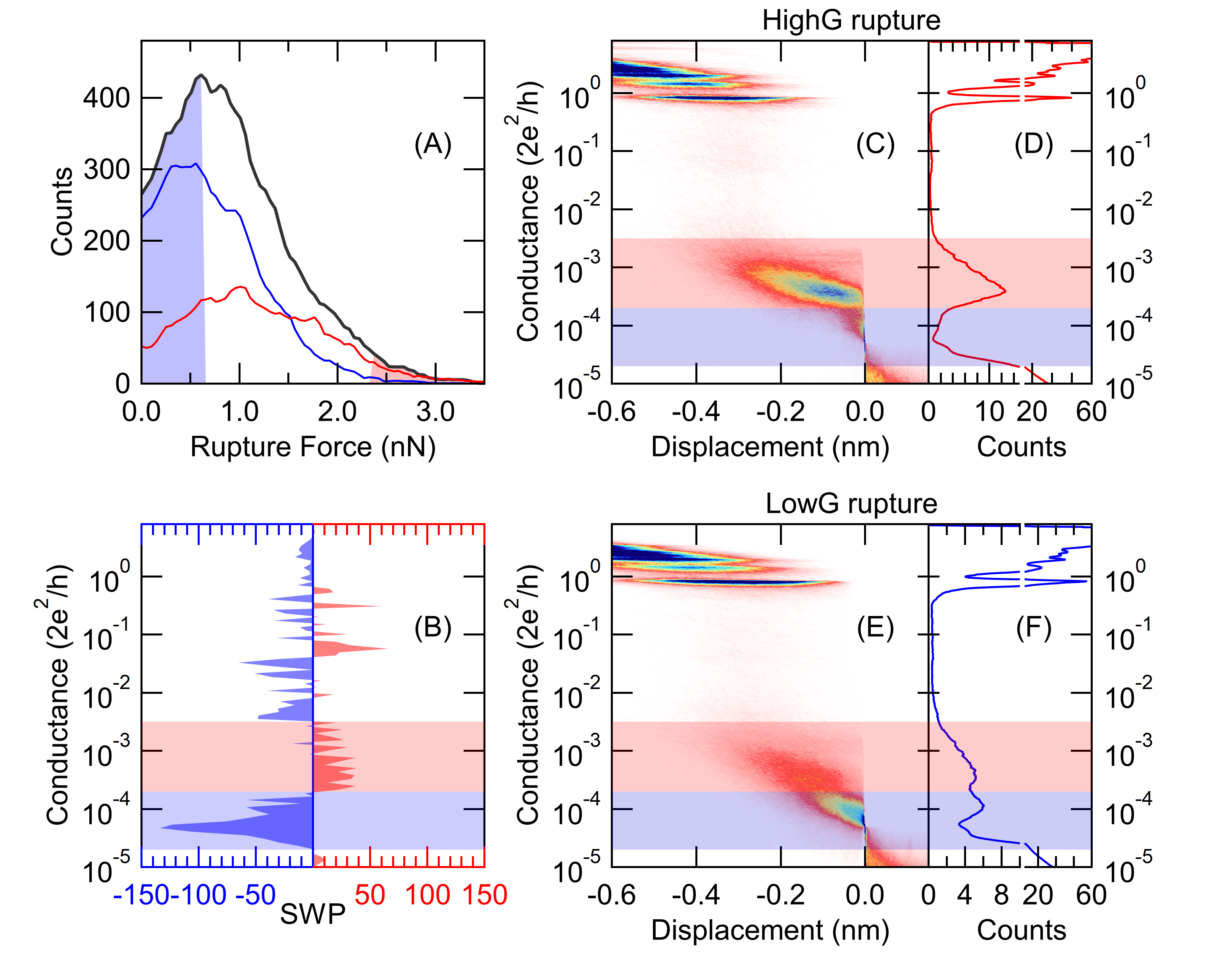}
\caption{\textit{
Unsupervised classification of room temperature molecular traces relying on training labels from auxiliary rupture force data. (A) Distribution of the measured rupture force values: all traces (black line), training traces with low/high rupture force (light blue/red area), resulting trace classes (red/blue lines for HighG rupture and LowG rupture traces, respectively). (B) SWP plot of the trained network. (C-F) Conductance histograms of the resulting trace classes: molecular junctions rupturing from highG (C, D) and lowG (E, F) conductance configuration. The Light red and blue regions reflect the relevant intervals of the SWP plot. In the 2D histograms the traces are aligned at $G_\textrm{ref}=5\cdot10^{-5}\,$G$_0$.}}
\label{Fig6}
\end{center}
\end{figure}

\emph{(iv) Unsupervised classification of the conductance traces using auxiliary force measurements as training labels.}
So far the automatically generated training labels were supported by the extreme values of the principal component projections. In the following we demonstrate, that additional measured quantities (like rupture force data) can also be used to select trace classes according to the distinct values of these auxiliary measured quantities. Such selected traces can be used as training sets for the neural network. This training might empower the network to recognize the typical motifs of the classes solely in the conductance data such that the auxiliary quantity is only required for the training, but not for the inference. To demonstrate this scheme we use the room temperature BP break junction measurements (Fig.~\ref{Fig1}A,B), which are also extended by force data. The black line in Fig.~\ref{Fig6}A shows the distribution of the rupture force values calculated for all traces i.e. the force that is required to rupture the molecular junction.\cite{Frei2011, Frei2012, Aradhya2012a} As two classes we take traces with outstandingly large/small rupture force. More specifically we take the traces, for which the rupture force is within the top/bottom $20\%$ of the entire force range of $0-3\,$nN (see the regions highlighted by light red/light blue in Fig.~\ref{Fig6}A, respectively). In this special case the top $20\%$ of the force range contains almost an order of magnitude less traces than the bottom $20\%$, so we use the former traces multiple times ($10 \times$) to ensure a balanced training set. 1D conductance histograms of these traces are used to train the network. The network trained in this manner classifies all the traces into the two classes demonstrated by the 2D and 1D histograms in Fig.~\ref{Fig6}C-F, respectively. It is clear that the traces in the first class (high rupture force) break from the highG configuration (\emph{HighG rupture traces}), whereas in the second class (low rupture force) the traces break from the lowG configuration (\emph{LowG rupture traces}).  These trace classes are in full agreement with the conclusions of Aradhya et al.\cite{Aradhya2012}, where this dataset was first analyzed. Again, we can state that our neural network based analysis could automatically deliver the relevant trace classes of the dataset.
It is interesting to point out, that the rupture force distributions of the resulting groups of traces show a significant overlap, meaning that it would not be possible to achieve this classification by simply setting a threshold for the rupture force value. However the neural network trained on the traces with extreme rupture force values was able to identify the important features in the conductance data and generalize this on the rest of the conductance traces. The SWP plot (Fig.~\ref{Fig6}B) shows that the most relevant part of the input data comes from the conductance range corresponding to the lowG configuration.

\section{Conclusions}
In conclusion, we have have analyzed the data of single-molecule break junction measurements exploiting neural networks, which are very efficient in feature recognition. The direct application of neural networks on break junction data is obviously restricted by the requirement of training data, which are already labeled according to the targeted classes. In break junction measurements, however, the relevant trace classes are not necessarily known in advance, the researchers seek well-suited algorithms, which help to identify the presence of distinct junction formation trajectories. To this end, we have developed a combined method, which automatically generates training data according to the extreme values of the principal component projections or some auxiliary measured quantities, and then the network captures the features of these characteristic traces, and generalizes its inference to the entire dataset. This method matched our personal judgment with $93\%$ accuracy in classifying molecular vs. tunneling traces. More importantly, we have demonstrated that our combined approach is able to recognize much less obvious, but highly relevant trace classes in a fully automated and unsupervised manner. Finally, we have demonstrated that the simplest possible double layer feed forward neural networks are not only enough for the proper classification, but the simple network structure also enables a direct insight to the decision making mechanism through the summed weight product plots.  We believe, that this unsupervised recognition of the relevant trace classes provides a fundamental support to final goal of understanding the physical mechanisms in the ultimate smallest conductors.

\section{Acknowledgements}
This work was supported by the BME-Nanonotechnology FIKP grant of EMMI (BME FIKP-NAT) and the NKFI K119797 grant. L.V. thanks the NSF-DMR 1807580 grant for support. The authors are thankful to Kasper P. Lauritzen and Gemma C. Solomon for useful discussions and for inspiration to this work, and to Sriharsha V. Aradhya and Michael Frei for the room temperature force measurement data. 

\section{Methods}
The low temperature measurements were performed by notched wire mechanically controllable break junctions using $100\,\mu$m diameter high purity gold wires which were fixed to a bending beam by two drops of stycast epoxy. The cross-section of the wire was reduced by a notch in the middle. The wire was broken in situ in the cryogenic vacuum using a coarse mechanical actuation. The conductance of the junction was recorded by applying a small ($100\,$mV) bias voltage and measuring the current with a logarithmic current amplifier.\cite{Meszaros2007} To record a statistical amount of conductance traces the junction was opened and closed many thousands of times with a piezo actuator. The molecules were evaporated to the junction in situ by heating a quartz tube inside the tungsten coil of a light bulb.\cite{Magyarkuti2016} The distance between the neighbor peaks in the $1\,$G$_0$ step length histogram\cite{Yanson1998} was used to calibrate the displacement of the junction. 

The room temperature measurements were performed with a scanning tunneling microscope break junction arrangement which was supplemented with an AFM cantilever to precisely measure the force as well.\cite{Aradhya2012} In this measurement, molecules were evaporated onto gold coated mica substrate. 

The one dimensional (1D) histograms were created for each trace by dividing the conductance axis to small ranges (bins) and calculating the number of measurement points in each conductance bin. Then single trace histograms are averaged to calculate the 1D histogram representing the entire dataset. 2D histograms are created by aligning each trace at the crossing of a given conductance level ($G_\textrm{ref}$) and performing binning along both the conductance and displacement axes. The neural network algorithm was implemented using the TensorFlow machine learning platform.\cite{Abadi2016}

\bibliographystyle{prsty}
\bibliography{main}

\end{document}